\documentclass[twocolumn]{revtex4-1}
\usepackage{graphicx}
\usepackage{dcolumn}
\usepackage{bm}
\usepackage{amssymb}
\usepackage{latexsym}
\usepackage{amsmath}
\usepackage{tensor}
\begin{document}
\title{Formalism for stochastic perturbations and analysis in relativistic
 stars}
\author{ Seema Satin }
\affiliation{Indian Institute for Science Education and Research,Pune India}
\email{seemasatin74@gmail.com}
\newcommand{\tG}{\tensor{G}}
\newcommand{\D}{\tensor{D}}
\newcommand{\N}{\tensor{N}}
\newcommand{\hT}{\hat{T}}
\newcommand{\C}{\mbox{Cov}}
\newcommand{\be}{\begin{equation}}
\newcommand{\ee}{\end{equation}}
\newcommand{\bea}{\begin{eqnarray}}
\newcommand{\eea}{\end{eqnarray}}
\newcommand{\G}{\hat{G}}
\newcommand{\e}{\epsilon}
\begin{abstract}
Perturbed Einstein's equations with a linear response relation and a 
stochastic source, applicable to a relativistic star model are worked out .
These perturbations which are stochastic in nature, are of significance for
 building a non-equilibrium statistical theory in connections with 
relativistic astrophysics. 
A fluctuation dissipation relation for  a spherically symmetric star
in its simplest form is obtained. The FD relation shows how the  
 random velocity fluctuations in the background of the
unperturbed star can  dissipate into Lagrangian displacement of fluid 
trajectories of the  dense matter. Interestingly in a simple way,  a 
constant (in time) coefficient of dissipation is obtained without a delta
correlated noise.  This formalism is also extended  for perturbed TOV equations
 which have a stochastic contribution, and show up in terms of the 
effective or root mean square pressure perturbations. Such 
 contributions can shed light on new ways of analysing the equation of state
 for dense matter.
 One may obtain  contributions of first and second order in the equation of
 state using this stochastic approach.
\end{abstract}
\maketitle
\section{Introduction}
 Perturbations in relativistic stars have been of significance
towards asteroseismology, and stability criteria in massive stars 
\cite{fried,kok1,jackson}.
The mode analysis of perturbations plays a central role in 
 gravitational wave detection \cite{allen,nils1,nils2}.  
 The theoretical developments for this are based on the perturbed Einstein's
 equations.
In this article  a part of the new basic formalism for relativistic stars and
their  perturbative analysis with a stochastic approach is given . This  is in
continuation with an upcoming theme of research \cite{seema1,seema2,seema3}, 
using the framework of  Langevin formalism  in general relativity. The open
directions in this regard deal with formulations and solutions for a classical 
Einstein-Langevin equation  for various astrophysical models. For
 considerations in general relativity, the Langevin noise  
has to be obtained  in an elaborate form and is model dependent.
Similar developments in the semiclassical theory are well 
established \cite{bei1} for semiclassical Einstein-Langevin equation. 
However the semiclassical developments follow  differently  in terms of
 formulations and applications due to the quantum fields and quantum stress
 tensors that define
the  Langevin  noise. The applications of the semiclassical stochastic gravity
 is towards early universe cosmology \cite{bei2} and black hole physics 
\cite{sukanya}.  We do not 
follow those lines, though the idea of a Langevin noise defined due to 
fluctuations in the matter fields is borrowed from the semiclassical
 counterpart. 
 In many other areas of physics it is easy to model the system
by assuming gaussian white noise or some coloured form of the noise 
term in the Langevin equation \cite{Chandra,risken} in a handwaving way.
One cannot do this for the framework which we are developing here. 
It is the spacetime metric
and the matter fields, that decide the form of the Langevin
term for the Einstein's equations. This has to be obtained elaborately, using 
the generalized  fluctuations
 of the respective stress-energy tensors. 
The generalized fluctuations as defined  in \cite{seema4} are an extension of
the random fluctuations in time to that of random fluctuations on a spacetime,
w.r.t. the spatial as well as temporal coordinates.
We  use the terminology, generalized fluctuations, in our theme
of research for the background noise in the unperturbed system. 
 The term perturbations is specifically used for the induced
shift in the trajectories of the fluid and the metric potentials. The 
background fluctuations in the fluid matter can be due to many reasons, they
can be thermal or non-thermal in nature.
For example in a perfect fluid having no, or negligible thermal effects,
 such generalized stochastic fluctuations can arise
either due to dynamical effects of fluid on small scale, or  as
remnants (coarse grained effects)  of quantum mechanical fluctuations in the
 bulk. These are 
expected to show up at  sub-hydro mesoscopic scales which lie below the 
macroscopic hydro scales but much above the quantum mesoscopic or microscopic
 scales.

In this article, we assume the radial velocity fluctuations in the background
spacetime as noise,  with a  near-equilibrium (dynamical) configuration of the
relativistic star.
It is shown how these may induce random perturbations in the system. The
radial perturbations are of significance towards stability properites of the
relativistic star \cite{fried,nils1}.
 In this article we also
 work out a fluctuation dissipation relation for a spherically symmetric 
configuartion in a simple way. The perturbed TOV relation obtained in the 
later section
is due to contributions from the generalised stochastic pressure perturbations 
which are induced due to the Langevin noise. In the concluding section we give 
the importance of developing such a framework and further directions for 
theoretical  formulations and investigations.
 
We begin by reviewing the Einstein-Langevin equation based on a linear 
response theory \cite{kubo} in the section below.
\section{The linear response relation for the perturbed Einstein's equation 
with a source term}
 A linear response relation for perturbations of relativistic stars has been
introduced in \cite{seema3}, which gives the following form of 
the Einstein-Langevin equation. 
\be \label{eq:ltreq}
 \delta G_{ab}[h;x) - 8 \pi \delta T_{ab}[h;x) - 8 \pi \int K(x-x')
\delta T_{ab}[\xi]; x')
= \tau_{ab}[g;x)
\ee 
where $h$ denotes the metric perturbations and $\xi$, the fluid perturbations.
The factor $K(x-x')$ is the response kernel and connects the metric
and fluid perturbations. 
In the above equation $\tau_{ab}[g;x) $ denotes the stochastic source inside 
 the dense matter of the relativistic star. 
We define $\tau_{ab}(x) = \delta_s T_{ab}(x) $, in terms of the generalized 
stochastic fluctuations of  matter fields of the massive astrophysical 
object.   
The linearized perturbed Einstein's equations are  covaraintly  conserved
 w.r.t the background metric $g_{ab}$, which
also implies $ \nabla_a \tau^{ab}(x) = 0 $.  
In case  the perturbations in the gravitating system are due to external
sources, then we assume $\tau_{ab}(x) = 0 $ in equation (\ref{eq:ltreq}),
 and  $K(x-x') = \delta^4(x-x')$. In this case  one obtains the
conventional perturbed Einstein's equation given by $\delta G_{ab}(x)
= 8 \pi \delta T_{ab}(x) $, which forms the base of asteroseismology. One
can  also consider a configuartion where few components of $\tau_{ab}(x) $ 
but not all, may be zero. We will see this more clearly in the example that we 
solve in this article. 
 We model with $\tau_{ab}(x)$ only the internal sources which may
 perturb the astrophysical body.  

In the next two sections we solve the  classical Einstein-Langevin 
equation for a spherically symmetric model of the star with 
non-thermal background velocity fluctuations in the fluid. The configuration
 is assumed to be near a static equilibrium state by the end of the 
collapse, and has matter fields described by a perfect fluid. 
\section{ The spherically symmetric model of relativistic star and
a stochastic source of induced perturbations}
For the spherically symmetric star in Schwarzchild coordinates,
\be
ds^2 = - e^{2 \nu(r)} dt^2 + e^{2 \lambda(r) } dr^2 + r^2 d \Omega^2
\ee
 while  the matter fields for perfect fluid are described by,  
\be
T_{ab} = (\epsilon+ p) u_a u_b + g_{ab} p
\ee
 the four-velocity is given by,
\be
u^a =  (e^{-\nu}, 0 , 0,0).
\ee
At a later time a radial velocity which maintains spherical symmetry 
can be introduced in the fluid given by, 
$ v = e^{\lambda - \nu} \dot{r}$, such that the four-velocity  has components
\be
u^a = \frac{1}{\sqrt{1-v^2}} (e^{-\nu}, e^{-\lambda} v , 0,0)
\ee
Accordingly, the components of field equation are given as
\bea
G^t_t = 8 \pi T^t_t & : & \nonumber \\
  e^{-2 \lambda} (\frac{1}{r^2} - \frac{2}{r}
\lambda ' )& - & \frac{1}{r^2} = - 8 \pi \epsilon \frac{1}{1-v^2} \label{eq:1}\\
G^r_r = 8 \pi T^r_r & : & \nonumber \\
e^{-2 \lambda}(\frac{1}{r^2} + \frac{2}{r} \nu' ) & - & \frac{1}{r^2} = 8
\pi ( \epsilon \frac{v^2}{1-v^2} + p ) \label{eq:2} \\
G^t_r = 8 \pi T^t_r & : & \nonumber \\
- \frac{2}{r} e^{-2 \nu} \dot{\lambda}  & = & 8 \pi e^{\lambda - \nu}
(\epsilon + p) \frac{v}{1- v^2} \label{eq:3}\\
e^{ 2 \lambda} G^\theta_\theta = 8 \pi e^{2 \lambda}  T^\theta_\theta
& : & \nonumber \\
\nu '' + \nu'^2 & - & \nu ' \lambda ' +  \frac{1}{r} (\nu ' - \lambda ' )
= 8 \pi e^{2 \lambda} p
\eea
Also we can easily obtain the relation,
\be
\nu' + \lambda' = 4 \pi (\epsilon + p ) e^{2 \lambda} p
\ee
from the above Einstein's equations. 
Perturbations of the above equations can be carried out in a heuristic way,
 which suffice out requirements. The perturbations in the fluid can be
intoduced using the radial velocity such that, 
\be
\delta v = e^{\lambda - \nu} \dot{\xi}
\ee
where $\xi \equiv \xi_r$ is the only non-zero fluid displacement vector.    
The terminology that we use here should be clarified as "perturbations "
meaning shift in trajectory of the fluid and related physical 
quantities in the system, while the generalized "fluctuations" is used
for the noise in the unperturbed background which we define in the
subsection below. 
\subsection{The source term with four-velocity fluctuations} 
We consider a model of the source term or generalized noise $\tau_{ab}(x)$ with
randomness over the spatial as well as temporal coordinates.
We will show that the perturbations are induced in the star due to the 
cummulative effect of these generalized fluctuations. 

The specific model of the noise or source that we consider can be given by
the single non-zero term of $\tau_{ab}(x) $ .
\be
\tau^t_r(r,t) = 8 \pi e^{(\lambda  - \nu)} (\epsilon + p) \delta_s v(t,r) 
\ee
where $\delta_s v(t,r)$ denotes the generalized fluctuations in the radial
velocity of the fluid. The 's' in $\delta_s v(t,r)$ denotes "source". 
The probability distribution of $P(\delta_s v (r,t)) $ 
is crucial to the nature of solutions of equation (\ref{eq:ltreq}). The
source is  Langevin  if the  distribution is gaussian, such that
 $<\tau_{ab}(x)> = 0 $, where $<....> $ denotes the
statistical average. The two point correlations 
 $<\tau_{ab}(x) \tau_{cd}(x') >_s = N_{abcd}(x,x') $ define the point
separated noise kernel. For the Langevin noise then, all the 
higher order correlations can be described in terms of the two point
noise kernel or correlations.  
\section{Solution of the Einstein Langevin equation} 
Given the model of noise in the previous subsection  and equation 
 (\ref{eq:1}) (\ref{eq:2}) and (\ref{eq:3}) for the
 spherically symmetric relativistic star, the perturbed
equations with the source described by  (\ref{eq:ltreq})  have
$t-t, r-r, t-r, \theta-\theta, \phi-\phi$ as the non-zero components.
We  require the following perturbed components of the E-L equation for a 
complete solution of the configuration discussed here,
\bea
& &  \delta G^t_t[h;x) - 8 \pi \delta T^t_t[h;x) - 8 \pi \delta T^t_t[\xi;x)
= 0  \label{eq:tt} \\
& & \delta G^r_r[h;x)  - 8 \pi \delta T^r_r[h;x) - 8 \pi \delta T^t_r[\xi;x)
= 0  \label{eq:rr} \\
& & \delta G^t_r[h;x) - 8 \pi \delta T^t_r [h;x) - 8 \pi \int K(x-x')
 \delta T^t_r[\xi; x')  dx' \nonumber\\
& & =  \tau^t_r[g,x) \label{eq:tr}
\eea
In the above, the components $t-t$ and $r-r$ do not contain a source term, 
since $\tau^t_t$ and $\tau^r_r $ are zero for the model of noise that
we use here. Thus, as discussed 
earlier, we assume a delta correlated response kernel 
$ K(x-x') = \delta^4 (x-x') $ for these two components of the E-L equation. 
To solve the above equations we will also need to specify the
form of the  Lagranigian displacement vector $\xi$, we assume it to be of
the form $\xi(r,t) = \tilde{\xi}(r) e^{\gamma_r t} $, where $\gamma_r$ is
complex valued. We choose a 
simple form of the response kernel $K(x-x') = K_1(t-t') \delta(r-r')$ 
for equation (\ref{eq:tr}) which gives, 
\begin{widetext}
\be
 \dot{\delta \lambda}(r,t) + (\nu' + \lambda') \int K(t-t') 
\dot{\xi}(r,t')  dt' = - e^{\lambda - \nu} (\lambda' + \nu') \delta_s v(r,t)
\ee
\end{widetext}
Taking $t-t' = \mathbf{T} $, we can bring the above in the form   
\be \label{eq:lamb}
\delta \lambda(r,t) = (\nu' + \lambda') K_1(\gamma_r) \xi(r,t) - 
e^{(\nu-\lambda)} (\nu' + \lambda') \int \delta_s v(r,t) dt
\ee
where $K_1(\gamma_r) $ is the Laplace transform of $K_1(t-t')$ and 
gives susceptibility of perturbations in the astrophysical configuration.
With this we will show a relation between the $\xi(r,t) $ and the noise
term.  
Using (\ref{eq:tt}) one can get the following relation
\bea \label{eq:xi2}
& & - \frac{2}{r} \delta \lambda' + \delta \lambda ( \frac{3 \lambda'}{r}
- \frac{2}{r^2} - \frac{\nu'}{r} ) = \frac{(\nu'+ \lambda')}{r} \xi' 
\nonumber \\
& & + \frac{(\nu'+ \lambda')}{2} ( \lambda' + \frac{2}{r} ) \xi + 8 \pi 
\epsilon' e^{2 \lambda} \xi 
\eea
From (\ref{eq:lamb}), we get 
\bea \label{eq:lamb1}
\delta \lambda'(r,t) & = & [(\nu' + \lambda') K_1(\gamma_r) \xi(r,t)]' 
- e^{\nu - \lambda} [({\nu'}^2 - {\lambda'}^2 ) \nonumber \\
& &  + (\nu'' + \lambda'') ] \int \delta_s v(r,t) dt 
\eea
in the above we assume $\delta_s' v(r,t) = 0 $.

Substitute equation (\ref{eq:lamb}) and (\ref{eq:lamb1}) in (\ref{eq:xi2})
in terms of  $\xi(r,t)$  we get, 
\be \label{eq:xi3}
g(r) \xi'(r,t) + f(r) \xi(r,t) = j(r) \int \delta_s v(r,t) dt
\ee
where
\bea
 g(r)& =& -\frac{(\lambda' + \nu')}{r} (2 K_1(\gamma_r) + 1) \nonumber  \\
f(r) &=& [(\nu''+ \lambda'')K_1(\gamma_r) + (\nu' + \lambda') \{K_1'(\gamma_r)
+ \nonumber \\
& & K_1(\gamma_r)\{\frac{3 \lambda'}{r} - \frac{2}{r^2} - \frac{\nu'}{r} \}
+ \lambda' + \frac{2}{r} \}] \nonumber \\ 
 j(r) &=& e^{\nu - \lambda} [ \lambda'' + \nu'' + (\nu' + \lambda') 
(\frac{3  \lambda'}{r} - \frac{2}{r^2} - \frac{\nu'}{r} ) \nonumber\\
& &  - ({\nu'}^2 - {\lambda'}^2)  \}] \nonumber  
\eea
The solution of (\ref{eq:xi3}) gives,
\be \label{eq:delxi}
\xi(r,t)  =  m_1(r) \int \{m_2(r') \int \delta_s v(r',t') dt' \} dr' 
\ee
where 
\bea
m_1(r) & = &  e^{-\int \frac{f(r')}{g(r')} dr' } \nonumber \\
m_2(r) & = &  e^{\int \frac{f(r')}{g(r')} dr'} \frac{j(r)}{g(r)} \nonumber
\eea
We can conclude  form the above that $\xi(r,t)$ is random in nature due to
being induced by $\delta_s v(r,t) $.  With the form it has $\tilde{\xi}(r) 
e^{\gamma_r t}$, which is random in $\tilde{\xi}(r)$ as well as
$\gamma_r$ can be random. The complex nature of the frequency $\gamma_r$ 
is important as, this will be used  for mode analysis of the stochastic
perturbations in future work.  
Using equation (\ref{eq:delxi}) for $\xi(r,t)$ in equation (\ref{eq:lamb}) 
we get,
\bea \label{eq:dellam}
\delta \lambda(r,t) & = & l_1(r)   \int \{  m_2(r')
 \int \delta_sv(r',t') dt' \} dr' - \nonumber \\
& & l_2(r) \int \delta_s v(r,t') dt'
\eea
where
\bea
l_1 & = & (\nu' + \lambda') K_1(\gamma_r) e^{- \int \frac{f(r')}{g(r')} dr'}
\nonumber \\
l_2 &= & e^{\nu-\lambda} (\nu' + \lambda')  \nonumber
\eea
From equation (\ref{eq:rr}) similarly one  can obtain,
\begin{widetext}
\bea
\delta \nu' & = & 2 \delta \lambda e^{-2 \lambda} ( \frac{1}{r^2} +
 \frac{2}{r} \nu') + 8 \pi \Gamma_1 p [ - \delta \lambda - 
 \frac{e^{-\lambda}}{r^2} [e^\lambda r^2 \xi]' - \xi p'
\eea
which on substituting for $\delta \lambda$ and $\xi$ gives,
\be \label{eq:delnu}
\delta \nu(r,t)  =  \int \int a_1(r') m_2(r'') \delta_s v(r'',t') 
dt' dr'' dr' + \int \int a_2(r') \delta v(r',t') dt' dr' 
\ee
where
\bea
a_1 & = & (\nu' + \lambda ') K_1(\gamma_r) [ 2 e^{-2 \lambda} (\frac{1}{r^2}
+ \frac{2}{r} \nu' ) - \Gamma_1 p]  - 8 \pi \Gamma_1 p (\lambda' + \frac{2}{r}) + 8 \pi \Gamma_1 p 
\frac{f(r)}{g(r)}
  \nonumber \\
a_2 & = & 8 \pi \Gamma_1 p \frac{j(r)}{g(r)} - e^{\nu-\lambda} (\nu' + \lambda')
\{ 2 e^{-2 \lambda }(\frac{1}{r^2} + \frac{2}{r} \nu' ) - \Gamma_1 p \}
\nonumber
\eea
\end{widetext}
The potentials $\delta \lambda(r,t) $ and $ \delta \nu(r,t)$ also have a random
nature due to being induced by $\delta_s v(r,t)$.
All the unperturbed variables in the above expressions are functions of $r$ 
only. Equations  (\ref{eq:delxi}), (\ref{eq:dellam})  and
(\ref{eq:delnu})  give  the main expressions for $\delta \lambda(r,t),
\xi(r,t)$ and $\delta \nu(r,t)$. These expressions show that the
 perturbations are  induced by the cummulative effect of the radial velocity
 fluctuations which are the background noise. These the expressions are  
meaningful only as statistical results, hence we need to take statistical
 averages for $\delta \lambda(t,r)$, $ \delta \nu(t,r)  $ and $ \xi(t,r)$,
and appropriately write them as 
\begin{widetext}
\be
\langle \xi(r,t) \rangle  =  m_1(r) \int \{m_2(r') \int \langle
 \delta_s v(r',t') \rangle dt' \} dr' = 0 
\ee
\be
\langle \delta \lambda(r,t)\rangle  =  l_1(r)   \int \{  m_2(r')
 \int \langle \delta_s v(r',t') \rangle  dt' \} dr' - l_2(r)
 \int \langle \delta_s v(r,t') \rangle dt' = 0 
\ee
\bea
\langle \delta \nu(r,t) \rangle  =   \int a_1(r') m_2(r'') 
\langle \delta_s v(r'',t') \rangle dt' dr'' dr' + 
\int  a_2(r') \langle \delta v(r',t') \rangle dt' dr'= 0 
\eea
We see that these are vanishing due to the Langevin property of the noise.
However the two point correlations of these perturbations are
non-vanishing and  read,
\bea 
\langle \xi(r_1,t_1) \xi(r_2,t_2) \rangle & = & m_1(r_1) m_1(r_2)  \int
\{m_2(r_1') m_2(r_2') \int  \langle \delta_s v (r_1',t_1') \delta_s
v(r_2',t_2') \rangle dt_1' dt_2' \} dr_1' dr_2'
\eea 
\bea
\langle \delta \lambda(r_1, t_1) \delta \lambda(r_2,t_2) \rangle & = & 
l_1(r_1) l_1(r_2) \int  \{ m_2(r_1') m_2(r_2') \int \int \langle \delta_s
v(r_1',t_1') \delta_s v(r_2',t_2') \rangle dt_1' dt_2' \} dr_1' dr_2' 
\nonumber \\
& & + l_2r_1) l_2(r_2)   \int \langle \delta_s v(r_1, t_1')
\delta_s v(r_2,t_2') \rangle dt_1' dt_2' - l_1(r_1) l_2(r_2) \nonumber \\
& & \int \{ m_2(r_1')
\int  \langle \delta_s v(r_1',t_1') \delta_s v(r_2,t_2')\rangle dt_1' dt_2'
\} dr_1'  - l_2(r_1) l_1(r_2) \int \{m_2(r_2') \nonumber \\
& & \int \langle \delta_s v(r,t_1')
\delta_s v(r_2',t_2') \rangle dt_1' dt_2' \} dr_2' 
\eea
\bea
\langle \delta \nu(r_1,t_1) \delta \nu(r_2,t_2) \rangle & = & \int
 a_1(r_1') a(r_2') m_2(r_1'') m_2(r_2'') \langle \delta_s v(r_1'',t_1')
\delta_s v(r_2'',t_2') \rangle dt_1' dt_2' dr_1'' dr_2'' dr_1' dr_2'
  \nonumber \\
& & + \int a_2(r_1') a_2(r_2') \langle \delta_s v(r_1',t_1') 
\delta_s v(r_2',t_2')\rangle dt_1' dt_2' dr_1' dr_2'  \nonumber \\
& & + \int a_1(r_1') a_2(r_1')  m_2(r_1'') \langle \delta_s v(r_1'',t_1')
\delta_s v(r_2',t_2') \rangle dt_1' dt_2' dr_1'' dr_1' dr_2' \nonumber \\
& & \int a_2(r_1') a_1(r_1') m_2(r_2'') \langle \delta_s v(r_1', t_1')
\delta_s v(r_2'',t_2') \rangle dt_1' dt_2' dr_1' dr_2'' dr_2' 
\eea
\end{widetext}
From the above one can easily take  the coincidence of points 
and obtain root mean square value for the generalized stochastic 
perturbations. It is the rms value that gives the strength of these
 perturbations  and fluctuations for
 a localized point in the spacetime. The 
point separated or two point correlations are important to probe the 
non-local and extended properties of the matter fields. These nonlocal
or point separated correlations of the induced perturbations
then carry the signature of the inherent nature of state of matter at 
mesoscopic scales.
In the mode analysis of these perturbations, the essential features 
at  mesoscopic scale would be reflected which will characterize  the state of
 matter and its composition. Such modes
 which we will work out in another article with a detailed analysis, can be
 appropriately called "stochastic modes" of the induced perturbations. 
\section{Fluctuation-Dissipation Relation}
The fluctuation dissipation relation is valid near the static equilibrium state
of the relativistic star. Taking equation (\ref{eq:tr}) once more into
consideration we can write,
\begin{widetext}
\be  \label{eq:FD1}
\dot{\delta \lambda(r,t)} - (\nu'(r) + \lambda'(r)) \int K_1(t-t')
 \dot{\xi}(r,t') dt'  = e^{\nu(r) - \lambda(r)}
(\nu'(r) + \lambda'(r)) \delta_s v(r,t)
\ee
\end{widetext}
where "." denotes derivative w.r.t time. 
In the above $\dot{\delta \lambda}(r,t) $, can be  considered negligible for
probing the local matter fields, as the metric perturbation $\delta \lambda$ 
itself is small, near the static equilibrium state. 
 The massive star configuration  can be considered to collapse from an initial
 state at  $t= - \infty $ to $t=0$.
 We consider the limit $t=0$ as the static equilibrium stage, near which
 we work out this relation.
 The near-static-equilibrium state can be interesting due to the 
adiabatic stochastic perturbations which characterize the dense matter at
sub-hydro mesoscopic scales.
Here we propose to use our formalism, to define  a non-thermal
adiabatic fluctuation-disspiation relation. From (\ref{eq:FD1})
assuming $\dot{\delta \lambda} = 0$ near $t=0$,  we obtain
\be
\delta_s v(r,t) = K_1(\gamma_r) \gamma_r e^{\lambda(r) - \nu(r) }\xi(r,t)
\ee
\begin{widetext}
The two point correlation gives,
\bea
<\delta_s v(r,t) \delta_s v^*(r',t')>& = & K_1(\gamma_r) K_1(\gamma_{r'})
<\gamma_{r} \gamma_{r'}^*> e^{\{(\lambda(r)+ \lambda(r')) -
 (\nu(r)+ \nu(r')) \} } <\xi(r,t) \xi^*(r',t')>  \nonumber \\
\N{^t_r^t_r}(r,t,r',t') & = & S(r,r') \D{^t_r^t_r}(t,r,t',r')
\label{eq:FD2}
\eea
where the dissipation kernel is given by the only non-zero component  
\be  
\D{^t_r^t_r}(r,t,r',t') =  e^{\{(\lambda(r)+ \lambda(r')) -
 (\nu(r)+ \nu(r')) \} } <\xi(r,t) \xi^*(r',t')> 
\ee
\end{widetext}
and the noise kernel by the only non-zero component
\be
 \N{^t_r^t_r}(r,t,r',t') = <\delta_s v(r,t) \delta_s v^*(r',t')>
\ee
 while  $  K_1(\gamma_r) K_1(\gamma_{r'}) <\gamma_{r} \gamma^*_{r'}> = S(r,r')$ is
the coefficient of dissipation.
 Equation  (\ref{eq:FD2}) gives the fluctuation-disspation relation, where
 the lhs defines the correlation of generalized fluctuations in the fluid.
This F-D relation shows that the velocity fluctuations near equilibrium
configuration of the star dissipate their energy into inducing the 
 the Lagrangian displacement of fluid trajectories  namely $\xi(r,t) $. 
 Thus one can view the adiabatic dissipative phenomena in the fluid as 
a result of  $\delta_s v(r,t) $ acting as seeds for the  
induced perturbations of the fluid. These seeds in the form of 
the velocity fluctuations can be due to various reasons, including 
the end of dynamically collapsing phase or some other internal mechanical 
phenomena which leaves its footprints on the matter. 
 Thus the mechanical microscopic effects coarse grained as
 $\delta_s v(r,t) $ can be seen to act as sources of perturbations in the 
system.  
We would be interested in the root mean square for (\ref{eq:FD2}), which
takes the form with, $S(r) = K_1(\gamma_r)K_1(\gamma*_r)
  |\gamma_{r}|^2  $. So that the rms value for the fluctuations 
can be written as
\be
\delta_s v(r,t)_{\mbox{rms}} = \sqrt{S(r)}  e^{\lambda(r) -\nu(r)}
\xi(r,t)_{\mbox{rms}}
\ee
As we see, $S(r)$ is independent of the $t$ variable, it gives a 
constant w.r.t time. This is interesting, since without considering
a delta correlated noise, we have been able to obtain a dissipation constant 
independent of time. 
 This should not be a surprise, as we work 
on a static background equilibrium state, such that the noise is 
defined in an ad-hoc way  on $g_{ab}(x) $ which is static. The dissipation coefficient
describes the background properties, hence it is a consistent result. 
\section{Perturbed TOV equations} 
The perturbed TOV equation gives the structure of the near equilibrium
relativistic star and  can be obtained by considering $\nabla \delta T^a_1
=0 $, which gives
\be
\delta p'= - \nu'(\delta p + \delta \epsilon) - e^{2(\lambda - \nu)}
(\epsilon + p ) \ddot{\xi}(r,t)
\ee
The above equation can be written as 
\begin{widetext}
\be \label{eq:tov1}
\delta p(r,t)  =  -\int \{ \nu'(r')(\delta p(r') + \delta \epsilon(r')) + 
e^{2(\lambda(r') - \nu(r'))} 
  (\epsilon(r') + p(r') ) \ddot{\xi}(r',t) \} dr'
\ee
\end{widetext}
This is stochastic in nature due to the Langevin formalism. 
Moreover it is the root mean square of these perturbations which gives the
 effective value and can be added to the background unperturbed pressure
for a new equilibrium state. 
Such an analysis can be suitable to study stability 
properties of the star, and also may shed light on the equation of state, when
used as an additive contribution to the regular TOV equation which is of the
form,
\be
 p(r) = - \int \nu'(r') (\epsilon(r') + p(r')) dr'
\ee 
For equation (\ref{eq:tov1}), taking the two point correlation and the 
coincidence limit,
\begin{widetext}
\bea  \label{eq:tov2}
\lim_{r_1,t_1 \rightarrow r_2,t_2}< \delta p(r_1,t_1) \delta p_(r_2,t_2)> 
& = & \lim_{r_1',t_1 \rightarrow r_2',t_2}
\int \int [\nu'(r_1')\nu'(r_2') <(\delta p(r_1',t_1) + \delta \epsilon(r_1,t_1)
)( \delta p(r_2',t_2) + \delta \epsilon(r_2',t_2) > \nonumber \\
& & + e^{2(\lambda(r_1') + \lambda(r_2') - \nu(r_1') - \nu(r_2') }
(\epsilon(r_1') + p(r_1') )(p(r_2') + \epsilon(r_2'))  
(\gamma_{r_1'}\gamma_{r_2'})^2 <\xi(r_1', t_1) \xi(r_2',t_2)> \nonumber \\
& & \nu'(r_1') \gamma_{r_2'} e^{2(\lambda(r_2') - \nu(r_2'))} 
<(\delta \epsilon(r_2',t_2) + \delta p(r_2',t_2)) \xi(r_2',t_2)> 
+  \gamma_{r_1'} \nu'(r_2') (\epsilon(r_1') + p(r_1')) \nonumber \\
& & <\xi(r_1',t_1) (\delta p(r_2',t_2) + \delta \epsilon(r_2',t_2))>] 
dr_1' dr_2'   
\eea
The root mean square then can be easily obtained  and put in as an additive
part in the regular TOV equation to get,
\be \label{eq:tov}
p(r) + (\delta p(r,t))_{rms} = - \int \nu'(r')(\epsilon(r') +p(r')) dr'
 + \sqrt{ \lim_{r_1,t_1 \rightarrow r_2,t_2} 
 <\delta p(r_1,t_1) \delta p(r_2,t_2)> } 
\ee 
\end{widetext}
where for the second term on the rhs of equation (\ref{eq:tov}), one has to 
evaluate the  rhs of equation (\ref{eq:tov2}). 
To solve this perturbed TOV equations one needs to consider an equations of
 state like $p(r) = K_2 \epsilon^{\Gamma}(r) $ and its perturbation
$\delta p(r,t) = K_2 \Gamma (\delta \epsilon (r,t))^{(\Gamma -1)}(r,t) $ . 
The quantity on the lhs of equation (\ref{eq:tov}) gives the new
near equilibrium dynamical  pressure in the fluid . 
Note that on the rhs of equation (\ref{eq:tov2}) the correlations between the
pressure and energy density perturbations with the Lagrangian displacement
vector are non vanishing.  One can also use this perturbed equation for
 finding out
the first and second order corrections, to the equation of state using the
rms value and coincidence of the two point correlation/variance, respectively.
\section{Concluding Remarks}
In this article we have discussed  the detailed form of the classical Einstein
Langevin equation  for induced perturbations, due 
to velocity fluctuations as the background source. The results obtained are
in closed analytical form. The solutions  show dependence on the susceptibilty
of the spacetime  perturbations. 
We have discussed the simple case of spherically symmetric spacetime in
order to establish the main ingredients for  the new theoretical base.
These new theoretical foundations have to be developed rigorously from the
 basics in general relativity and the framework of Langevin approach  
in a classical mesoscopic scenario . 
More realistic and wider range of configurations of the relativisitic stars 
in a perturbatve study  for non-radial cases can be based on the framework 
presented here. Most of the realistic configurations will need 
numerical solutions for the Einstein-Langevin equations,  which will
 be carried out later.  The radial non-rotating case of the relativistic stars 
is just the beginning of the research theme that is to follow.
 Our further goal in this theme of research is to develop 
the same analysis for non- radial perturbations. We will also work on
perturbations  arising  due to other matter fields like the electromagnetic
 field coupled to a relativistic star in near future.
 In this article, we have also 
discussed  a linear response relation for perturbations in relativistic stars
from the first principles. 
We show in a very simple way, how a perturbed  TOV equation turns
up with the stochastic formulation, and give the basic expressions for the 
same. This can be used to study new modes of random oscillations in the
relativistic stars which may be named as "stochastic modes". We
are  interested in characterizing such stochastic modes for non-radial
random oscillations in future.  It is through such formulations as 
presented here, that we intend 
to develop non-equilibrium and equilibrium theory for mesoscopic scales
in the dense matter stars. One can analyse 
extended and non local properties of the dense matter, through two point or
higher correlations of the induced perturbations. Hence this is the first step 
as the theoretical base ( along with work done in and developments in 
\cite{seema1,seema2,seema3,seema4,seema5} )
into  investigations that are to follow for the dense matter in gravitating
 bodies. The main interest in this direction of research lies in
analysis of properties of  the dense matter at mesoscopic sub-hydro scales
for local as well as extended structure in the relativistic stars. 


\begin{thebibliography}{00}
\bibitem{fried} John.L.Friedman., The Astro.Jour, 200 (1975) 204-220.
\bibitem{kok1} K.D.Kokkotas and J.Ruoff. A\&A, 366 No. 2 (2001) 565-572.
\bibitem{jackson} J.C.Jackson., Mon.Not.R.Astron.Soc.276, (1995)
965-970. 
\bibitem{allen} Gabrielle Allen. et al., Phys Rev D 58 (1998) 124012.
\bibitem{nils1} Nils Andersson et al., Mon.Not. Roy.Astron. Soc. 274 (1995)
1039.
\bibitem{nils2} Nils Andersson, K.D.Kokkotas and Bernard F.Schutz. Mon.Not.R.
Astron. Soc, (1996) 1230-1234.
\bibitem{seema1} Seema Satin., Gen.Rel and Grav. 50: 97 (2018) .
\bibitem{seema2} Seema Satin., Gen.Rel and Grav. 51:52 (2019).
\bibitem{seema3} Seema Satin., (under review) arXiv:2110.01837v1[gr-qc] 2021.
\bibitem{bei1} Bei Lok Hu and Enric Verdaguer., Liv.Rev.Rel  11:3 (2008).
\bibitem{bei2} B.L.Hu and A.Matacz., Phys Rev D 51, (1995) 1577.
\bibitem{sukanya} Sukanya Sinha, Alpan Raval and B.L.Hu. Found. of Phys. 33
(1):37-64 (2003).
\bibitem{Chandra} S. Chandrashekhar., Rev.Mod. Phys 15,1 (1943).
\bibitem{risken}H.Risken.,The Fokker Planck Equation. second edition. Springer
Verlag. (1989)
\bibitem{seema4} Seema Satin., CQG 39 : 9 (2022) 095004. 
\bibitem{kubo} Toda.M, Kubo.R,Saito.N. Statistical Physics !: Equilibrium
Statistical Mechanics. (1983). 
\bibitem{seema5} Seema Satin.,Phys.RevD 100, (2019) 044032.
\end{thebibliography}
\end{document}